\documentclass[aps,prd,onecolumn,showpacs,showkeys,amsmath,amssymb]{revtex4}
\usepackage{amsfonts}
\usepackage{amsmath}
\usepackage{graphicx}
\usepackage{subfigure}
\usepackage{dcolumn}
\usepackage{bm}
\usepackage{booktabs}
\usepackage{ulem}
\usepackage[usenames,dvipsnames]{color}

\begin{document}
\title{QCD sum rule Study of the $d^*(2380)$}
\author{Hua-Xing Chen}
\affiliation{
School of Physics and Nuclear Energy Engineering and International Research Center for Nuclei and Particles in the Cosmos, Beihang University, Beijing 100191, China
}
\author{Er-Liang Cui}
\affiliation{
School of Physics and Nuclear Energy Engineering and International Research Center for Nuclei and Particles in the Cosmos, Beihang University, Beijing 100191, China
}
\author{Wei Chen}
\email{wec053@mail.usask.ca}
\affiliation{
Department of Physics and Engineering Physics, University of Saskatchewan, Saskatoon, SK, S7N 5E2, Canada
}
\author{T. G. Steele}
\email{tom.steele@usask.ca}
\affiliation{
Department of Physics and Engineering Physics, University of Saskatchewan, Saskatoon, SK, S7N 5E2, Canada
}
\author{Shi-Lin Zhu}
\email{zhusl@pku.edu.cn}
\affiliation{
School of Physics and State Key Laboratory of Nuclear Physics and Technology, Peking University, Beijing 100871, China \\
Collaborative Innovation Center of Quantum Matter, Beijing 100871, China \\
Center of High Energy Physics, Peking University, Beijing 100871, China
}
\begin{abstract}
We systematically construct $I(J^P)=0(3^+)$ six-quark local interpolating currents without derivative operators. We discuss the best choice of operator, and select three $\Delta$-$\Delta$ like operators to perform QCD sum rule analyses to calculate the mass of the $d^*(2380)$. The mass extracted from this analysis is $M_{d^*} = 2.4\pm0.2$ GeV, consistent with the $d^*(2380)$ mass observed by the WASA detector at COSY. We also obtain a sum-rule lower mass bound $M_{d^*} > 2.25$ GeV. We also consider the effect of mixing of singlet dibaryon fields with the same quantum numbers, and perform the QCD sum rule analysis of the mixed interpolating current and extract the mass of the $d^*(2380)$ and its lower mass bound. With optimized mixing parameters, we find that the mixed current does not change the numerical result significantly.
\end{abstract}
\pacs{12.39.Mk, 12.38.Lg}
\keywords{Dibaryon, QCD sum rule}
\maketitle
\pagenumbering{arabic}
%
%===============================================================
%===============================================================
\section{Introduction}
%===============================================================
%===============================================================
%

Recently, the WASA detector at COSY confirmed their early observations of the $d^*(2380)$~\cite{Bashkanov:2008ih,Adlarson:2011bh,Adlarson:2012fe,Adlarson:2014pxj,Adlarson:2014tcn}: to explore the nature of the ABC effect~\cite{Booth:1961zz}, they measured the polarized $\vec n p$ scattering, and their partial-wave analysis exhibited a resonance pole at $(2380\pm10 - i40\pm5)$ MeV with quantum numbers $I(J^P) = 0(3^+)$ \cite{Adlarson:2014ozl}.

The $d^*(2380)$ may be a $\Delta \Delta$ dibaryon or a six-quark state, first predicted on the basis of $SU(6)$ symmetry by Dyson and Xuang in 1964~\cite{dyson}. Since then, many theoretical efforts have been made to study this state. In Ref.~\cite{Goldman:1989zj} it was first denoted by $d^*(2380)$, and in Ref.~\cite{Oka:1980ax} Oka and Yazaki suggested its existence by investigating the one-gluon-exchange interaction between quarks. In Refs.~\cite{Zhang:1997ny,Yuan:1999pg,Huang:2014kja,Huang:2014aca,Huang:2013nba} various authors calculated its mass and decay width.  The Ref.~\cite{Huang:2014kja} calculation, where hidden color configurations are included, is in particularly good agreement with the Ref.~\cite{Adlarson:2014ozl} experimental finding. In Refs.~\cite{Gal:2013dca,Gal:2014zia} Gal and Garcilazo did a three-body study and they also obtained a mass consistent with the experimental value. In Ref.~\cite{Bashkanov:2013cla} the authors suggested that it has a ``hidden-color'' six-quark configuration. Recently, an alternative explanation of the $d^*(2380)$ was suggested by Bugg~\cite{Bugg:2013wxa}. The effect of the $d^*(2380)$ was discussed in Refs.~\cite{Faldt:2011zv,Albaladejo:2013sya} by analysing the WASA experiments (see also \cite{harvey,Flambaum:2007mj} for related studies). Besides the $d^*(2380)$, the H-dibaryon candidate proposed by Jaffe~\cite{Jaffe:1976yi} has also received considerable attention.

In this paper we study the $d^*(2380)$ using QCD sum rule techniques. We systematically construct $I(J^P)=0(3^+)$ six-quark local interpolating currents without derivative operators.
We present arguments for choosing three $\Delta-\Delta$ like operators, and perform QCD sum rule analyses to calculate the mass of the $d^*(2380)$. The extracted mass is consistent with the Ref.~\cite{Adlarson:2014ozl} experimental value.
We then consider mixing of these operators and perform the QCD sum rule analysis for this mixing interpolating current. The numerical results show that this mixed current does not change our previous conclusions significantly.
This paper is organized as follows. In Sec.~\ref{sec:fields}, we construct the interpolating currents for $d^*(2380)$ and select three $\Delta$-$\Delta$ like operators to perform QCD sum rule analyses in Sec.~\ref{sec:sumrule}. We study the mixing interpolating current in Sec.~\ref{sec:mixing}. A brief summary is given in Sec.~\ref{sec:summary}.

%
%===============================================================
%===============================================================
\section{$d^*(2380)$ Interpolating Fields}
\label{sec:fields}
%===============================================================
%===============================================================
%
In this section we construct six-quark local interpolating currents for $d^*(2380)$ without derivative operators. Local interpolating fields for dibaryons consisting of six quarks can be generally described as
\begin{eqnarray}
J(x) \sim \Gamma_{abcdef} \Gamma^{ABCDEF} \left( q^{aT}_A(x) \mathbb{C} \Gamma_1 q^b_B (x) \right) \left( q^{cT}_C(x) \mathbb{C} \Gamma_2 q^d_D (x) \right)  \left( q^{eT}_E(x) \mathbb{C} \Gamma_3 q^f_F (x) \right) \, ,
\label{eq:dibaryonfield}
\end{eqnarray}
where $a \cdots f$ are color indices, $A \cdots F$ are flavor indices, and $\Gamma_{abcdef}$ and $\Gamma^{ABCDEF}$ are used to denote their internal color and flavor structures, respectively; $q_{A}(x)=(u(x)\, ,d(x)\, ,s(x))$ is the flavor triplet quark field at location $x$; the superscript $T$ represents the transpose of the Dirac indices only;  and $\mathbb{C}$ is the charge-conjugation operator. In this paper we also use the following notations: $\lambda_N$ ($N=1\cdots8$) are the Gell-Mann matrices, $\epsilon^{ABC}$ is the totally anti-symmetric tensor, and $S_P^{ABC}$ ($P=1\cdots10$) are the normalized totally symmetric matrices.

Although the internal structure of the dibaryon can be very complicated, some simplifications are possible. First we investigate the color matrix $\Gamma_{abcdef}$. The dibaryon is a color singlet, and there are five possible combinations to compose a color-singlet object using six quarks:
\begin{eqnarray}
\mathbf{3}^6 = [ \mathbf{3} \otimes \mathbf{3} \otimes \mathbf{3} ]^2 &=& [ \mathbf{1} \oplus \mathbf{8}_A \oplus \mathbf{8}_B \oplus \mathbf{10} ]^2
\\ \nonumber &\rightarrow& \mathbf{1} \otimes \mathbf{1} = \mathbf{1} ~~~~~~~~~~~~~~ (a) \, ,
\\ \nonumber &\rightarrow& \mathbf{8}_A \otimes \mathbf{8}_A = \mathbf{1} ~~~~~~~~~~ (b) \, ,
\\ \nonumber &\rightarrow& \mathbf{8}_A \otimes \mathbf{8}_B = \mathbf{1} ~~~~~~~~~~ (c) \, ,
\\ \nonumber &\rightarrow& \mathbf{8}_B \otimes \mathbf{8}_A = \mathbf{1} ~~~~~~~~~~ (d) \, ,
\\ \nonumber &\rightarrow& \mathbf{8}_B \otimes \mathbf{8}_B = \mathbf{1} ~~~~~~~~~~ (e) \, .
\end{eqnarray}
The corresponding color matrices are
\begin{eqnarray}
\Gamma^{(a)}_{abcdef} &=& \epsilon_{abc} \epsilon_{def} \, ,
\\ \nonumber \Gamma^{(b)}_{abcdef} &=& \epsilon_{abg} \lambda^N_{gc} \epsilon_{deh} \lambda^N_{hf} = 2 \epsilon_{abf} \epsilon_{dec} - {\frac{2}{3}} \epsilon_{abc} \epsilon_{def} \, ,
\\ \nonumber \Gamma^{(c)}_{abcdef} &=& \epsilon_{abg} \lambda^N_{gc} \epsilon_{dhf} \lambda^N_{he} = 2 \epsilon_{abe} \epsilon_{dcf} - {\frac{2}{3}} \epsilon_{abc} \epsilon_{def} \, ,
\\ \nonumber \Gamma^{(d)}_{abcdef} &=& \epsilon_{agc} \lambda^N_{gb} \epsilon_{deh} \lambda^N_{hf} = 2 \epsilon_{afc} \epsilon_{deb} - {\frac{2}{3}} \epsilon_{abc} \epsilon_{def} \, ,
\\ \nonumber \Gamma^{(e)}_{abcdef} &=& \epsilon_{agc} \lambda^N_{gb} \epsilon_{dhf} \lambda^N_{he} = 2 \epsilon_{aec} \epsilon_{dbf} - {\frac{2}{3}} \epsilon_{abc} \epsilon_{def} \, .
\end{eqnarray}
We note there is another color structure $\mathbf{8}_C$ ($\epsilon^{gbc} \lambda_{ga}^N$) for color-singlet baryon fields, but it is related to $\mathbf{8}_A$ and $\mathbf{8}_B$ through the following relation:
\begin{eqnarray}
\epsilon^{abg} \lambda_{gc}^N + \epsilon^{bcg} \lambda_{ga}^N + \epsilon^{cag} \lambda_{gb}^N = 0 \, ,
\label{eq:Jacobi}
\end{eqnarray}
and therefore the color structure can always be simplified as combinations of two $\epsilon$ tensors. Moreover, these two $\epsilon$ tensors can be transformed to be $\epsilon_{abc} \epsilon_{def}$ via Fierz transformations. For example,
\begin{eqnarray}
&& \epsilon_{abf} \epsilon_{dec} \Gamma^{ABCDEF} \left( q^{aT}_A(x) \mathbb{C} \Gamma_1 q^b_B (x) \right) \left( q^{cT}_C(x) \mathbb{C} \Gamma_2 q^d_D (x) \right)  \left( q^{eT}_E(x) \mathbb{C} \Gamma_3 q^f_F (x) \right)
\label{eq:color1}
\\ \nonumber &\rightarrow& \epsilon_{abf} \epsilon_{dec} \Gamma^{ABCDEF} \sum_i \left( q^{aT}_A(x) \mathbb{C} \Gamma_{1,i} q^b_B (x) \right) \left( q^{fT}_F(x) \mathbb{C} \Gamma_{2,i} q^d_D (x) \right)  \left( q^{eT}_E(x) \mathbb{C} \Gamma_{3,i} q^c_C (x) \right)
\\ \nonumber &=& \epsilon_{abc} \epsilon_{def} \Gamma^{ABFDEC} \sum_i \left( q^{aT}_A(x) \mathbb{C} \Gamma_{1,i} q^b_B (x) \right) \left( q^{cT}_C(x) \mathbb{C} \Gamma_{2,i} q^d_D (x) \right)  \left( q^{eT}_E(x) \mathbb{C} \Gamma_{3,i} q^f_F (x) \right) \, ,
\end{eqnarray}
which can be further simplified to:
\begin{eqnarray}
(\ref{eq:color1}) &\rightarrow& \epsilon_{abc} \epsilon_{def} \sum_{i,j} \Gamma^{ABC}_j\Gamma^{DEF}_j \left( q^{aT}_A(x) \mathbb{C} \Gamma_{1,i} q^b_B (x) \right) \left( q^{cT}_C(x) \mathbb{C} \Gamma_{2,i} q^d_D (x) \right)  \left( q^{eT}_E(x) \mathbb{C} \Gamma_{3,i} q^f_F (x) \right)
\label{eq:color2}
\\ \nonumber &\rightarrow& \sum_{i,j} \mathcal{B}_{1,i,j}^T C \Gamma^\prime_{2,i} \mathcal{B}_{2,i,j} \, .
\end{eqnarray}
Here we use $\mathcal{B}_{1,i,j}$ and $\mathcal{B}_{2,i,j}$ to denote the two color-singlet baryon fields with flavor matrices $\Gamma^{ABC}_j$ and $\Gamma^{DEF}_j$, respectively. Eq.~\eqref{eq:color2} means that we can always transform a dibaryon local interpolating field into a combination of two color-singlet baryon fields by using Fierz transformations.

The local color-singlet baryon fields without derivatives have been systematically studied and classified in Ref.~\cite{Chen:2008qv}, where we found that there are altogether six independent baryon fields, including one flavor-singlet, three flavor-octet and two flavor-decuplet baryon fields. Among these fields, three of them do not have any free Lorentz indices:
\begin{eqnarray}
\Lambda_1 &=& \epsilon_{abc} \epsilon^{ABC} (q_A^{aT} C q_B^b) \gamma_5 q_C^c \, ,
\label{eq:baryon1}
\\ N^N_1 &=& \epsilon_{abc} \epsilon^{ABD} \lambda_{DC}^N (q_A^{aT} C q_B^b) \gamma_5 q_C^c \, ,
\label{eq:baryon2}
\\ N^N_2 &=& \epsilon_{abc} \epsilon^{ABD} \lambda_{DC}^N (q_A^{aT} C \gamma_5 q_B^b) q_C^c \, ,
\label{eq:baryon3}
\end{eqnarray}
two of them have one free Lorentz index:
\begin{eqnarray}
N^N_{3\mu} &=& P_{3/2}^{\mu\nu} N^{\prime N}_{3\nu} = N^{\prime N}_{3\mu} + {\frac{1}{4}} \gamma_\mu \gamma_5 (N^N_1 - N^N_2) \, ,
\label{eq:baryon4}
\\ \Delta^P_{1\mu} &=& \epsilon_{abc} S_P^{ABC} (\bar q_A^{aT} C \gamma_\mu q_B^b) q_C^c \, ,
\label{eq:baryon5}
\end{eqnarray}
where
\begin{eqnarray}
\nonumber N^{\prime N}_{3\mu} &=& \epsilon_{abc} \epsilon^{ABD} \lambda_{DC}^N (q_A^{aT} C \gamma_\mu \gamma_5 q_B^b) \gamma_5 q_C^c \, ,
\\ \nonumber P_{3/2}^{\mu\nu} &=& g^{\mu\nu} - {\frac{1}{4}}\gamma^\mu\gamma^\nu \, ,
\end{eqnarray}
and only one of them has two anti-symmetric free Lorentz indices:
\begin{eqnarray}
\Delta^P_{2\mu\nu} &=& P_{3/2}^{\mu\nu\alpha\beta} \Delta^{\prime P}_{2\alpha\beta} = \Delta^{\prime P}_{2\mu\nu} - {\frac{i}{2}} \gamma_\mu \gamma_5 \Delta^P_{1\nu} + {\frac{i}{2}} \gamma_\nu \gamma_5 \Delta^P_{1\mu} \, ,
\label{eq:baryon6}
\end{eqnarray}
where
\begin{eqnarray}
\nonumber \Delta^{\prime P}_{2\mu\nu} &=& \epsilon_{abc} S_P^{ABC} (q_A^{aT} C \sigma_{\mu\nu} q_B^b) \gamma_5 q_C^c \, ,
\\ \nonumber P_{3/2}^{\mu\nu\alpha\beta} &=& (g^{\mu\alpha}g^{\nu\beta} - {\frac{1}{2}}g^{\nu\beta}\gamma^{\mu}\gamma^\alpha + {\frac{1}{2}}g^{\mu\beta}\gamma^\nu\gamma^\alpha + {\frac{1}{6}}\sigma^{\mu\nu}\sigma^{\alpha\beta})\, .
\end{eqnarray}
We note that they all have a positive parity, except $\Delta^P_{2\mu\nu}$. The baryon fields with a negative parity can be obtained simply by inserting an extra $\gamma_5$.

We can now use these color-singlet baryon fields to construct the dibaryon interpolating fields. The $0(3^+)$ dibaryon field without derivatives can be generally written as
\begin{eqnarray}
J_{\alpha_1\alpha_2\alpha_3;\beta_1\cdots\beta_j} = \mathcal{S} [J^\prime_{\alpha_1\alpha_2\alpha_3;\beta_1\cdots\beta_j}] \, ,
\end{eqnarray}
where $\mathcal{S}$ denotes symmetrization and subtracting the trace terms in the sets $(\alpha_1\alpha_2\alpha_3)$, and $J^\prime_{\alpha_1\alpha_2\alpha_3;\beta_1\cdots\beta_j}$ result in  dibaryon interpolating fields containing $0(3^+)$ components. We note that these operators can contain more than three free Lorentz indices. A familiar example is the electromagnetic field $F_{\mu\nu}$ which has two antisymmetric Lorentz indices but has spin only 1. To construct such a dibaryon field: (a) the matrix between two baryon fields, $\Gamma^\prime_{2,i}$ of Eq.~(\ref{eq:color2}), should contain either one free Lorentz index $\gamma_\mu$, or two anti-symmetric Lorentz indices $\sigma_{\mu\nu}$; (b)additionally, we need two extra symmetric Lorentz indices. We find the following four currents containing $0(3^+)$ components:
\begin{eqnarray}
(N^{N}_{3\alpha_1})^T C \gamma_{\alpha_3} N^{M}_{3\alpha_2} \, , (\Delta^{P}_{1\alpha_1})^T C \gamma_{\alpha_3} \Delta^{Q}_{1\alpha_2} \, ,  (\Delta^{P}_{2\alpha_1\beta_1})^T C \gamma_{\alpha_3} \Delta^{Q}_{2\alpha_2\beta_2} \, , (\Delta^{P}_{2\alpha_1\beta_1})^T C \sigma_{\alpha_3\beta_3} \Delta^{Q}_{1\alpha_2} \, ,
\label{eq:currents1}
\end{eqnarray}
and the following four currents containing $0(3^\pm)$ components, i.e., both $0(3^+)$ and $0(3^-)$ components:
\begin{eqnarray}
(N^{N}_{3\alpha_1})^T C \sigma_{\alpha_3\beta_3} N^{M}_{3\alpha_2} \, , (\Delta^{P}_{1\alpha_1})^T C \sigma_{\alpha_3\beta_3} \Delta^{Q}_{1\alpha_2} \, , (\Delta^{P}_{2\alpha_1\beta_1})^T C \gamma_{\alpha_3} \Delta^{Q}_{1\alpha_2} \, , (\Delta^{P}_{2\alpha_1\beta_1})^T C \sigma_{\alpha_3\beta_3} \Delta^{Q}_{2\alpha_2\beta_2} \, .
\label{eq:currents2}
\end{eqnarray}
%In this paper we shall choose only the $\Delta-\Delta$ like operators to perform QCD sum rule analyses.
In Ref.~\cite{Chen:2012ex} all these baryon fields were used to perform QCD sum rule analyses. The results showed that only sum rules from $\Delta^P_{1\mu}$ are reasonable. Furthermore, the mass sum rules from $N^N_{3\mu}$ are unphysical (the spectral density is negative in the physical region) while those from $\Delta^P_{2\mu\nu}$ are too trival to give reliable results. Hence, $(\Delta^{P}_{1\alpha_1})^T C \gamma_{\alpha_3} \Delta^{Q}_{1\alpha_2}$ and $(\Delta^{P}_{1\alpha_1})^T C \sigma_{\alpha_3\beta_3} \Delta^{Q}_{1\alpha_2}$ may be good candidates for QCD sum rule studies. However, the latter current contains both $3^+$ and $3^-$ components, so we shall use the former one for the $d^*(2380)$. Considering isospin symmetry, we therefore choose the following $\Delta-\Delta$ like dibaryon interpolating field to perform QCD sum rule analyses in the next section:
\begin{eqnarray}
J_1^{\alpha_1\alpha_2\alpha_3} = \mathcal{S} [(\Delta^{++}_{1\alpha_1})^T C \gamma_{\alpha_3} \Delta^{-}_{1\alpha_2}] = \mathcal{S} [\epsilon_{abc} \epsilon_{def} (u^{aT} C \gamma_{\alpha_1} u^b) u^{cT} C \gamma_{\alpha_3} d^f (d^{dT} C \gamma_{\alpha_2} d^e)] \, .
\label{eq:deltadelta1}
\end{eqnarray}
We note that the exact $0(3^+)$ current is $\mathcal{S} [(\Delta^{++}_{1\alpha_1})^T C \gamma_{\alpha_3} \Delta^{-}_{1\alpha_2} - (\Delta^{+}_{1\alpha_1})^T C \gamma_{\alpha_3} \Delta^{0}_{1\alpha_2}]$. However, we shall also investigate the other two $\Delta-\Delta$ like fields to ensure that our conclusions are not significantly influenced by the choice of interpolating current:
\begin{eqnarray}
J_2^{\alpha_1\alpha_2\alpha_3} &=& - \mathcal{S} [(\Delta^{\prime ++}_{2\alpha_1\beta})^T C \gamma_{\alpha_3} \Delta^{\prime -}_{2\alpha_2\beta}] = \mathcal{S} [\epsilon_{abc} \epsilon_{def} (u^{aT} C \sigma_{\alpha_1\beta} u^b) u^{cT} C \gamma_{\alpha_3} d^f (d^{dT} C \sigma_{\alpha_2\beta} d^e)] \, ,
\label{eq:deltadelta2}
\\ J_3^{\alpha_1\alpha_2\alpha_3} &=& - \mathcal{S} [(\Delta^{\prime ++}_{2\alpha_1\beta})^T C \sigma_{\alpha_3\beta} \Delta^{-}_{1\alpha_2}] = \mathcal{S} [\epsilon_{abc} \epsilon_{def} (u^{aT} C \sigma_{\alpha_1\beta} u^b) u^{cT} C \sigma_{\alpha_3\beta} d^f (d^{dT} C \gamma_{\alpha_2} d^e)] \, .
\label{eq:deltadelta3}
\end{eqnarray}
By doing this we shall have investigated all the local $3^+$ dibaryon currents of color-singlet-color-singlet $\Delta-\Delta$ type. However, we note that there are still many other dibaryon fields with $I(J^P)=0(3^+)$, such as the $p$-$n$ type current $(N^{N}_{3\alpha_1})^T C \gamma_{\alpha_3} N^{M}_{3\alpha_2}$ and many other color-octet-color-octet currents and non-local currents. However, our primary interest is exploration of the $\Delta-\Delta$ scenarios, and hence these additional currents are beyond the scope of this work.

We assume $J_a^{\alpha_1\alpha_2\alpha_3}$ ($a=1,2,3$) couples to the $d^*(2380)$ with $I(J^P)=0(3^+)$ through
\begin{eqnarray}
\langle 0| J_a^{\alpha_1 \alpha_2 \alpha_3} | 0(3^+) \rangle = f_a \eta_{\alpha_1\alpha_2\alpha_3} \, ,
\end{eqnarray}
where $f_a$ is the decay constant; $\eta_{\alpha_1\alpha_2\alpha_3}$ is the symmetric and traceless polarization tensor, which has the following properties at the leading order (only counting $g_{\mu\nu}$):
\begin{eqnarray}
\eta_{\alpha_1\alpha_2\alpha_3} \eta^*_{\beta_1\beta_2\beta_3} = \mathcal{S}^\prime [g_{\alpha_1\beta_1} g_{\alpha_2\beta_2} g_{\alpha_3\beta_3}] \, ,
\end{eqnarray}
where $\mathcal{S}^\prime$ denotes symmetrization and subtracting the trace terms in the sets $(\alpha_1 \alpha_2 \alpha_3)$ and $(\beta_1 \beta_2 \beta_3)$.

\section{QCD Sum Rule Analysis}
\label{sec:sumrule}

QCD sum rule techniques have proven to be a powerful and successful non-perturbative method over the past few decades~\cite{Shifman:1978bx,Reinders:1984sr}. In sum rule analyses, we consider the following two-point correlation function:
%
%%%%%%%%%%%%%%%%%%%%%%%%%%%%%%%%%%%%%%%%%%%%%%%%%%%%%%%%%%%%%%%%%%%%%%%%%%%%%%
\begin{eqnarray}
\Pi_{\alpha_1\alpha_2\alpha_3, \beta_1 \beta_2 \beta_3}(q^2) &=& i \int d^4x e^{iqx} \langle 0 | T J_{\alpha_1\alpha_2\alpha_3}(x) J_{\beta_1 \beta_2 \beta_3}^\dagger (0) | 0 \rangle
\label{def:pi}
\\ \nonumber &=& (-1)^J \mathcal{S}^\prime [g_{\alpha_1\beta_1} g_{\alpha_2\beta_2} g_{\alpha_3\beta_3}] \Pi(q^2) \, ,
\end{eqnarray}
%%%%%%%%%%%%%%%%%%%%%%%%%%%%%%%%%%%%%%%%%%%%%%%%%%%%%%%%%%%%%%%%%%%%%%%%%%%%%%
%
and compute $\Pi(q^2)$ in the QCD operator product expansion (OPE) up to certain order in the expansion. The result is then matched with a hadronic parametrization to extract information about hadron properties. At the hadron level, we express the correlation function in the form of the dispersion relation with a spectral function:
%
%%%%%%%%%%%%%%%%%%%%%%%%%%%%%%%%%%%%%%%%%%%%%%%%%%%%%%%%%%%%%%%%%%%%%%%%%%%%%%
\begin{equation}
\Pi(q^2)={\frac{1}{\pi}}\int^\infty_{s_<}\frac{{\rm Im} \Pi(s)}{s-q^2-i\varepsilon}ds \, ,
\label{eq:disper}
\end{equation}
%%%%%%%%%%%%%%%%%%%%%%%%%%%%%%%%%%%%%%%%%%%%%%%%%%%%%%%%%%%%%%%%%%%%%%%%%%%%%%
%
where the integration starts from the physical threshold. The imaginary part of the two-point correlation function is
%
%%%%%%%%%%%%%%%%%%%%%%%%%%%%%%%%%%%%%%%%%%%%%%%%%%%%%%%%%%%%%%%%%%%%%%%%%%%%%%
\begin{eqnarray}
{\rm Im} \Pi(s) & \equiv & \pi \sum_n\delta(s-M^2_n)\langle 0|\eta|n\rangle\langle n|{\eta^\dagger}|0\rangle \, .
\label{eq:rho}
\end{eqnarray}
%%%%%%%%%%%%%%%%%%%%%%%%%%%%%%%%%%%%%%%%%%%%%%%%%%%%%%%%%%%%%%%%%%%%%%%%%%%%%%
%
As usual, we adopt a parametrization of one pole dominance for the ground state $d^*(2380)$ and a continuum contribution. The sum rule analysis is then performed after the Borel transformation of the two correlation function expressions (\ref{def:pi}) and (\ref{eq:disper})
%
%%%%%%%%%%%%%%%%%%%%%%%%%%%%%%%%%%%%%%%%%%%%%%%%%%%%%%%%%%%%%%%%%%%%%%%%%%%%%%
\begin{equation}
\Pi^{(all)}(M_B^2)\equiv\mathcal{B}_{M_B^2}\Pi(p^2) = {\frac{1}{\pi}} \int^\infty_{s_<} e^{-s/M_B^2} {\rm Im} \Pi(s) ds \, .
\label{eq:borel}
\end{equation}
%%%%%%%%%%%%%%%%%%%%%%%%%%%%%%%%%%%%%%%%%%%%%%%%%%%%%%%%%%%%%%%%%%%%%%%%%%%%%%
%
Assuming the contribution from the continuum states can be approximated well by the OPE spectral density above a threshold value $s_0$ (duality),
we arrive at the sum rule relation for the $d^*(2380)$. The sum rule for the first current $J_1^{\alpha_1\alpha_2\alpha_3}$ is
%
%%%%%%%%%%%%%%%%%%%%%%%%%%%%%%%%%%%%%%%%%%%%%%%%%%%%%%%%%%%%%%%%%%%%%%%%%%%%%%
\begin{eqnarray}
\label{eq:sumrule1}
f_1^2 e^{-M_{d^*}^2/M_B^2} &=& \Pi_1(s_0, M_B^2)
\\ \nonumber &=& \int^{s_0}_{0} e^{-s/M_B^2} ds \times \Big (
\frac{9}{8! 8! 44 \pi ^{10}} s^7 - \frac{343 \langle g_s^2 GG \rangle}{ 8! 8! 16 \pi^{10}} s^5 + \frac{\langle \bar q q \rangle^2}{ 4! 4! 14 \pi^6} s^4 + \frac{2\langle \bar q q \rangle \langle g_s \bar q \sigma G q \rangle}{ 4! 4! 3 \pi^6} s^3
\\ \nonumber && + \frac{9 \langle g_s \bar q \sigma G q \rangle^2}{ 4! 4! 10 \pi^6} s^2 - \frac{7 \langle g_s^2 GG \rangle \langle \bar q q \rangle^2}{4! 4! 20 \pi^6} s^2 +\frac{8 \langle \bar q q \rangle^4}{27 \pi ^2} s - \frac{23 \langle g_s^2 GG \rangle \langle \bar q q \rangle \langle g_s \bar q \sigma G q \rangle}{ 4! 4! 24 \pi^6} s
\\ \nonumber && + \frac{4\langle \bar q q \rangle^3 \langle g_s \bar q \sigma G q \rangle}{9 \pi^2} - \frac{3\langle g_s^2 GG \rangle \langle g_s \bar q \sigma G q \rangle^2}{ 4! 4! 16 \pi^6}
\Big ) - \frac{\langle g_s^2 GG \rangle \langle \bar q q \rangle^4}{162 \pi ^2} \, ,
\end{eqnarray}
and the one using the second current $J_2^{\alpha_1\alpha_2\alpha_3}$ is
\begin{eqnarray}
\label{eq:sumrule2}
f^2_2 e^{-M_{d^*}^2/M_B^2} &=& \Pi_2(s_0, M_B^2)
\\ \nonumber &=& \int^{s_0}_{0} e^{-s/M_B^2} ds \times \Big (
\frac{45}{8! 8! 88 \pi ^{10}} s^7 - \frac{2989 \langle g_s^2 GG \rangle}{ 8! 8! 64 \pi^{10}} s^5 + \frac{13\langle \bar q q \rangle^2}{ 4! 4! 224 \pi^6} s^4 + \frac{13 \langle \bar q q \rangle \langle g_s \bar q \sigma G q \rangle}{ 4! 4! 24 \pi^6} s^3
\\ \nonumber && + \frac{117 \langle g_s \bar q \sigma G q \rangle^2}{ 4! 4! 160 \pi^6} s^2 - \frac{\langle g_s^2 GG \rangle \langle \bar q q \rangle^2}{4! 4! 5 \pi^6} s^2 +\frac{13 \langle \bar q q \rangle^4}{54 \pi ^2} s - \frac{25 \langle g_s^2 GG \rangle \langle \bar q q \rangle \langle g_s \bar q \sigma G q \rangle}{ 4! 4! 48 \pi^6} s
\\ \nonumber && + \frac{13\langle \bar q q \rangle^3 \langle g_s \bar q \sigma G q \rangle}{36 \pi^2} - \frac{3\langle g_s^2 GG \rangle \langle g_s \bar q \sigma G q \rangle^2}{ 4! 4! 32 \pi^6}
\Big ) - \frac{\langle g_s^2 GG \rangle \langle \bar q q \rangle^4}{648 \pi ^2} \, .
\end{eqnarray}
%%%%%%%%%%%%%%%%%%%%%%%%%%%%%%%%%%%%%%%%%%%%%%%%%%%%%%%%%%%%%%%%%%%%%%%%%%%%%%
%
The sum rule for the third current $J_3^{\alpha_1\alpha_2\alpha_3}$ is identical to Eq.~(\ref{eq:sumrule1}) obtained for the first current. This may suggest that $J_1^{\alpha_1\alpha_2\alpha_3}$ and $J_3^{\alpha_1\alpha_2\alpha_3}$ can be related to each other, or the equivalence of the sum-rules may only exist at leading-order. Moreover, we shall see in the following discussions that the results obtained by using $J_1^{\alpha_1\alpha_2\alpha_3}$ and $J_2^{\alpha_1\alpha_2\alpha_3}$ are also quite similar. We discuss the results obtained from the first current $J_1^{\alpha_1\alpha_2\alpha_3}$ in detail, and simply show the results obtained from $J_2^{\alpha_1\alpha_2\alpha_3}$.

In Eqs.~(\ref{eq:sumrule1}) and (\ref{eq:sumrule2}) the chirally-suppressed contributions from current quark masses are neglected because they are numerically insignificant.
Moreover, the terms containing quark-related condensates in OPE are found to be significantly larger than those containing gluon-related condensates. In our calculation, we consider only leading order contributions from the two-gluon condensate ($\langle g_s^2 GG \rangle$). We would like to note that we have used the {\it Mathematica}  $FeynCalc$ package~\cite{feyncalc} to calculate the OPE. We have calculated the correlation function up to dimension 16 to make sure that the convergence of Eqs.~(\ref{eq:sumrule1}) and (\ref{eq:sumrule2}) is sufficient for a reliable QCD sum rule. To study this convergence, we use the following values for various condensates~\cite{Agashe:2014kda,Yang:1993bp,Gimenez:2005nt,Jamin:2002ev,Ioffe:2002be,Ovchinnikov:1988gk,colangelo,Hwang:1994vp,Narison:2002pw}:
%
%%%%%%%%%%%%%%%%%%%%%%%%%%%%%%%%%%%%%%%%%%%%%%%%%%%%%%%%%%%%%%%%%%%%%%%%%%%%%%
\begin{eqnarray}
\nonumber &&\langle\bar qq \rangle=-(0.240 \pm 0.010)^3 \mbox{ GeV}^3\, ,
\\
&&\langle g_s^2GG\rangle =(0.48\pm 0.14) \mbox{ GeV}^4\, ,
\label{condensates}
\\
\nonumber && \langle g_s\bar q\sigma G
q\rangle=-M_0^2\times\langle\bar qq\rangle\, ,
\\
\nonumber &&M_0^2=(0.8\pm0.2)\mbox{ GeV}^2\, ,
\end{eqnarray}
%%%%%%%%%%%%%%%%%%%%%%%%%%%%%%%%%%%%%%%%%%%%%%%%%%%%%%%%%%%%%%%%%%%%%%%%%%%%%%
%
and numerically show $\Pi_1(s_0, M_B^2)$ for $s_0 = \infty$:
\begin{eqnarray}
\Pi_1(\infty, M_B^2) &=& 6.8 \times 10^{-12} M_B^{16} - 8.0 \times 10^{-12} M_B^{12} + 5.9 \times 10^{-10} M_B^{10} - 1.1 \times 10^{-9} M_B^8
\label{25}
\\ \nonumber && + 2.8 \times 10^{-10} M_B^6 + 1.2 \times 10^{-9} M_B^4 - 1.3 \times 10^{-9} M_B^2 - 1.1 \times 10^{-11} \, .
\end{eqnarray}
%This equation tells us that when $M_B^2$ is around 1 GeV$^2$, we need to calculate the OPE at least up to dimension 16. However, the sum rule (\ref{eq:sumrule}) is negative and thus %unphysical at this point for $s_0 = \infty$.
Eq.~\eqref{25} shows that the OPE convergence improves with increasing $M_B$.
%At this point the sum rule (\ref{eq:sumrule}) is negative in a larger region $0 < M_B^2 <$ 2.2 GeV$^2$.
Therefore, we will work at the Borel mass region $M_B^2\geq$2.2 GeV$^2$ in our analysis to ensure that the perturbative contribution is the dominant contribution.
%to make sure the positivities of the two-point correlation function and the spectral density.
In this region, the dimension six and dimension eight condensates become significant to satisfy the OPE convergence.
%Hence, we need to use a  Borel mass at least larger than 2.2 GeV$^2$. Now the Dim=6 and Dim=8 terms become significant, making the convergence of OPE acceptable. \tcomm{This is a really %confusing discussion.  Could you not simplify this in some way?  Maybe just state that the OPE satisfies positivity and converges well for Borel scales above 2.2?}
In the following, we shall calculate the mass of the $d^*(2380)$ with $s_0$ around 6 GeV$^2$.

Another criterion of QCD sum rule is the pole contribution
%
%%%%%%%%%%%%%%%%%%%%%%%%%%%%%%%%%%%%%%%%%%%%%%%%%%%%%%%%%%%%%%%%%%%%%%%%%%%%%%
\begin{equation}
\label{eq_pole}
\mbox{Pole contribution} \equiv \frac{ \Pi_1(\omega_c, T) }{ \Pi_1( \infty , T) } \, .
\end{equation}
%%%%%%%%%%%%%%%%%%%%%%%%%%%%%%%%%%%%%%%%%%%%%%%%%%%%%%%%%%%%%%%%%%%%%%%%%%%%%%
%
However, it is only about 0.0002 when $M_B^2 \sim 5$ GeV$^2$ and $s_0 \sim 6$ GeV$^2$. We can not increase it to be larger than 0.01 in a wide region of 3 GeV$^2 <M_B^2< 15$ GeV$^2$ and 5 GeV$^2 <s_0< 7$ GeV$^2$. This is similar to other sum-rule analyses with multi-quark states, the pole contribution is small because of the large powers of $s$ in the spectral function.
In the following analyses, we find reasonable regions for the Borel mass $M_B$ and the threshold value $s_0$, in which the final results are stable against these two free parameters.

Our final expression for the mass of the $d^*(2380)$ is
%
%%%%%%%%%%%%%%%%%%%%%%%%%%%%%%%%%%%%%%%%%%%%%%%%%%%%%%%%%%%%%%%%%%%%%%%%%%%%%%
\begin{equation}
M^2_{d^*} = \frac{\frac{\partial}{\partial(-1/M_B^2)}\Pi_1(s_0, M_B^2)}{\Pi_1(s_0, M_B^2)} \, .
\end{equation}
%%%%%%%%%%%%%%%%%%%%%%%%%%%%%%%%%%%%%%%%%%%%%%%%%%%%%%%%%%%%%%%%%%%%%%%%%%%%%%

\begin{figure}[hbt]
\begin{center}
\scalebox{0.6}{\includegraphics{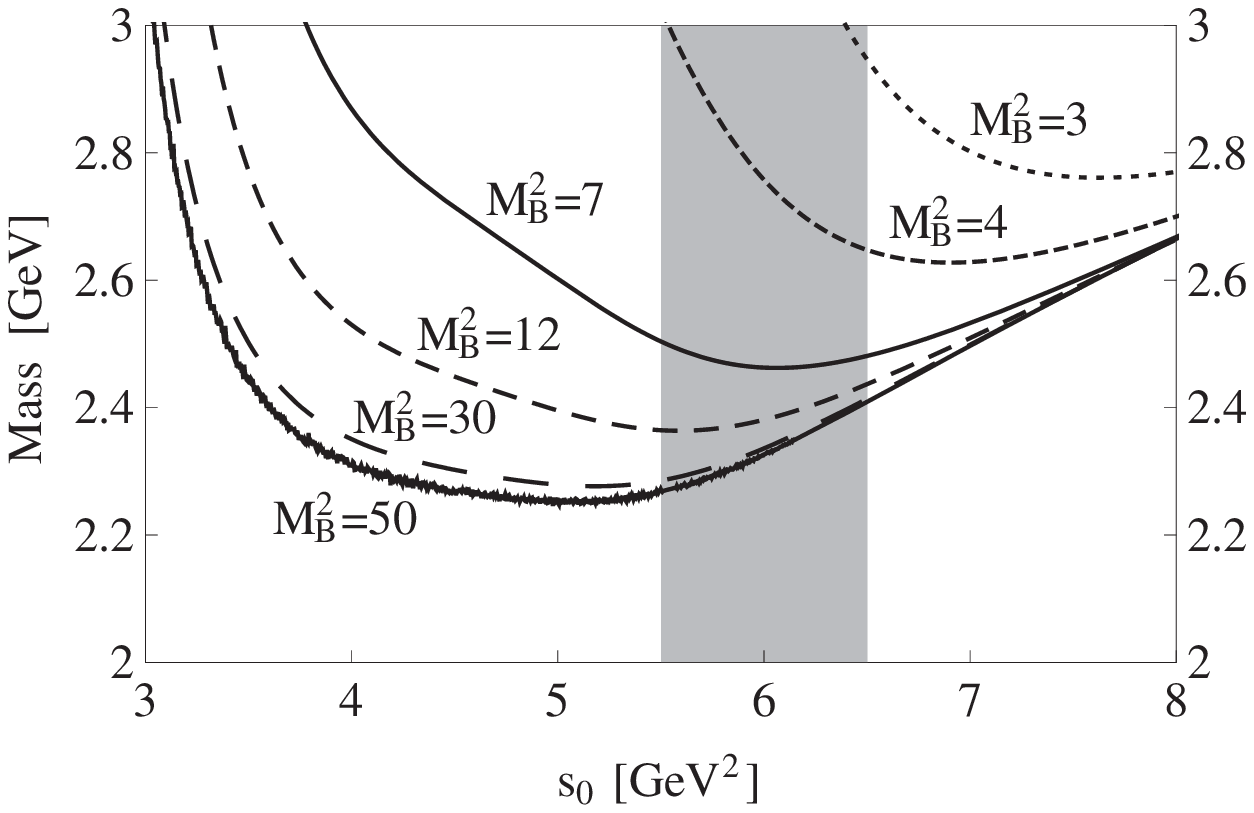}}
\scalebox{0.6}{\includegraphics{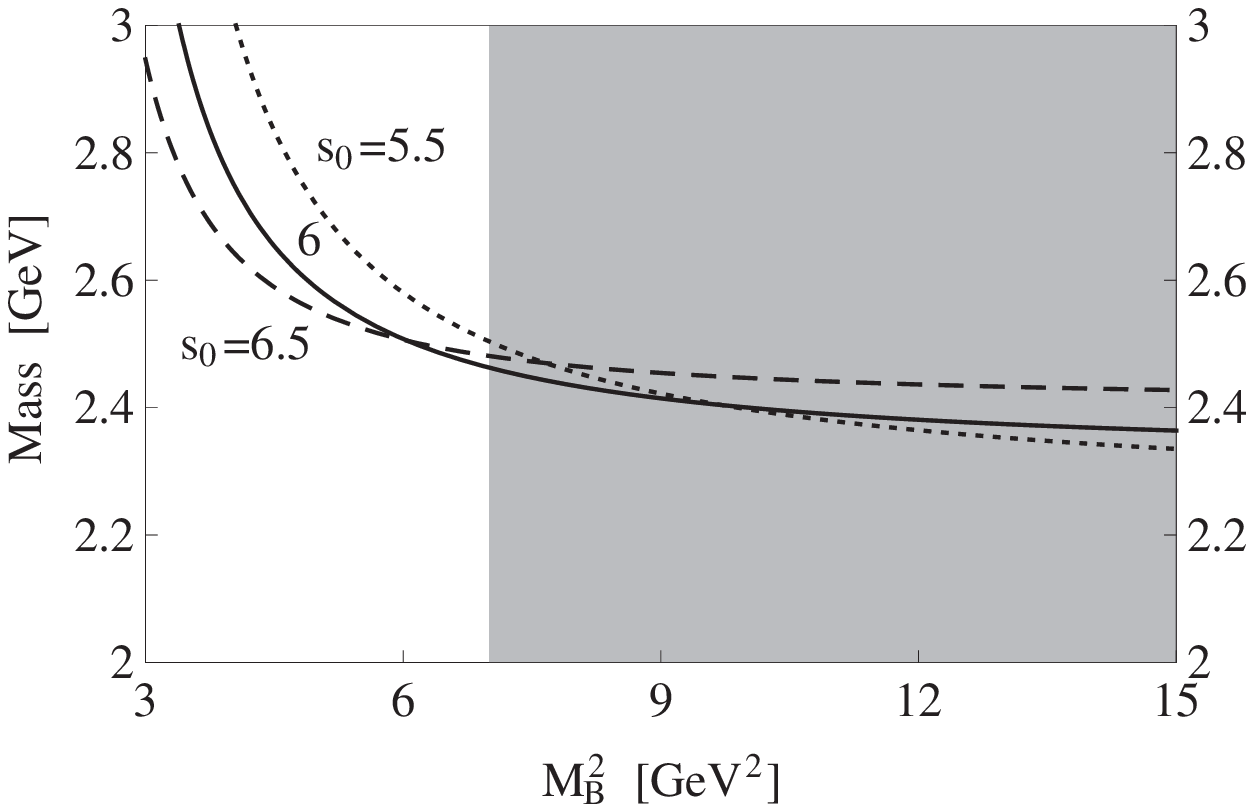}}
\centerline{\hspace{0.10in} {(a)} \hspace{2.85in}{ (b)}}
\caption{The variation of $M_{d^*}$ with respect to the threshold value $s_0$ (left) for $M_B^2 = 3,4,7,12,30,50$ GeV$^2$, and the Borel mass $M_B$ (right) for $s_0 = 5.5$ (dotted), $6.0$ (solid), and $6.5$ GeV$^2$ (dashed). The current $J_1^{\alpha_1\alpha_2\alpha_3}$ is used. The curves start to merge when $M_B^2$ is around 50 GeV$^2$, so we can not obtain a mass below 2.25 GeV.}
\label{fig:current1}
\end{center}
\end{figure}

In the Fig.~\ref{fig:current1}a we first show the variations of $M_{d^*}$ with respect to the threshold value $s_0$, in a large region  3 GeV$^2 < s_0 < 8$ GeV$^2$ and $M_B^2 = 3,4,7,12,30,50$ GeV$^2$. It shows that the obtained mass decreases monotonically and quickly with $M_B^2$ from 3 GeV$^2$ to 7 GeV$^2$. However, the dependence of our results on $M_B^2$ becomes much weaker as $M_B^2$ continues increasing from 7 GeV$^2$. All these curves have a mass minimum in a similar region when $s_0$ is around 5.5 GeV$^2$. Hence, the $s_0$ dependence of the mass prediction is also weak at this point. Accordingly, we use the following parameters as our working region:
\begin{eqnarray}
7 \mbox{ GeV}^2 < M_B^2 < 15 \mbox{ GeV}^2 \, , 5.5 \mbox{ GeV}^2 < s_0 < 6.5 \mbox{ GeV}^2 \, ,
\label{eq:region}
\end{eqnarray}
and obtain the mass of the $d^*(2380)$ to be
\begin{eqnarray}
M_{d^*} = 2.4\pm0.2 \mbox{ GeV} \, ,
\end{eqnarray}
where the error $\pm0.2$ GeV originates from the uncertainties in the Borel mass $M_B$, continuum threshold $s_0$ and the various condensates
shown in Eqs.~(\ref{condensates}).
The dominant error source in our analysis is the uncertainty of quark condensate.
We also evaluate the decay constant $f_1$ to be:
\begin{eqnarray}
f_1 = (5.5\pm2.7) \times 10^{-5} \mbox{ GeV}^8 \, ,
\end{eqnarray}
which gives the coupling between the interpolating operator and the physical state (the uncertainty in the coupling is based upon the same sources considered for the mass). This value of the coupling is needed to calculate hadronic decay width of the
physical state in the three-point function sum rules (see, e.g., Ref.~\cite{2013-Dias-p16004-16004}).
Moreover, we find that the curves start to fluctuate merge when $M_B^2$ is larger than about 50 GeV$^2$, which means that the variation of the
hadron mass with the Borel parameter $M_B^2$ is very small in this region.
Therefore, we find a lower bound on the $d^*(2380)$ mass of $M_{d^*} > 2.25$ GeV. For completeness, we also show the variations of $M_{d^*}$ with respect to the Borel mass $M_B^2$ in the  Fig.~\ref{fig:current1}b, again in a large region 3 GeV$^2 < M_B^2 < 15$ GeV$^2$ and for $s_0 = 5.5,6.0,6.5$ GeV$^2$, again illustrating that the dependence of our results on $M_B^2$ becomes much weaker when $M_B^2$ is larger than 7 GeV$^2$.

\begin{figure}[hbt]
\begin{center}
\scalebox{0.6}{\includegraphics{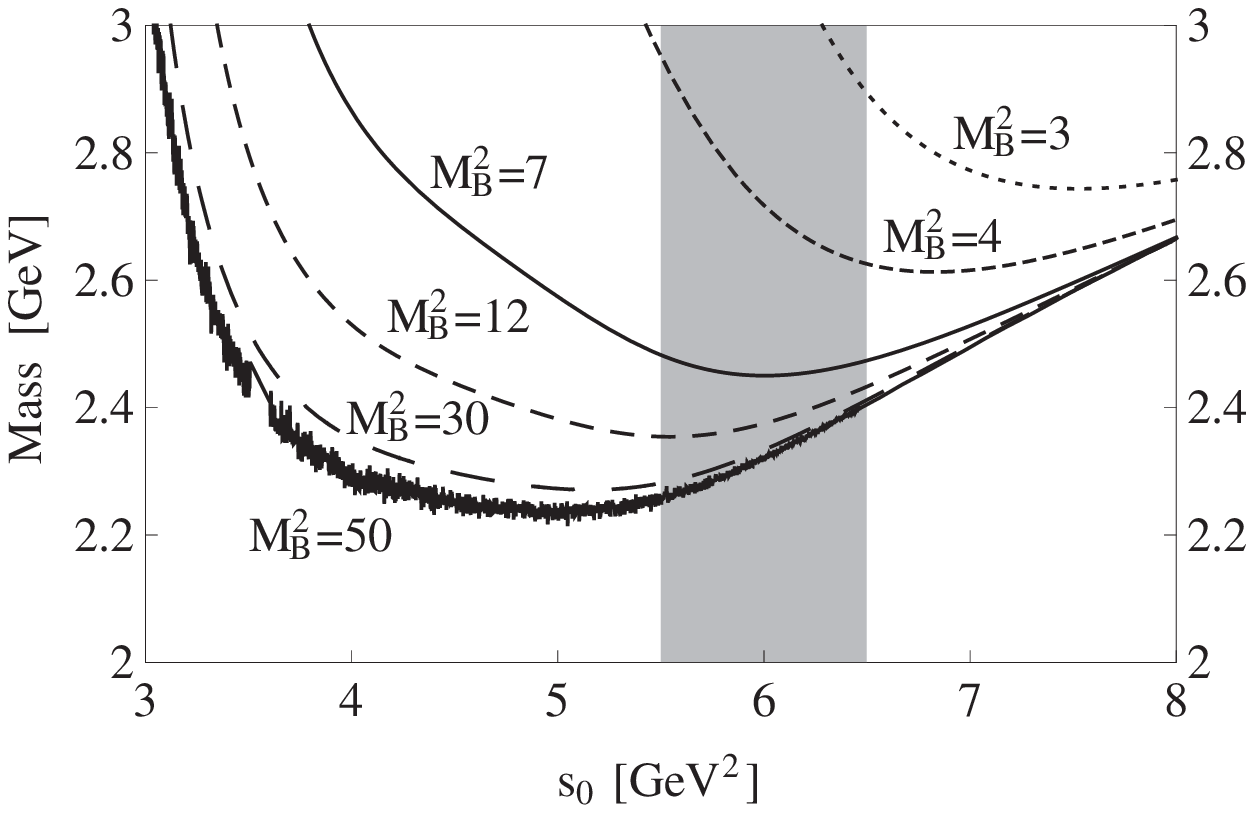}}
\scalebox{0.6}{\includegraphics{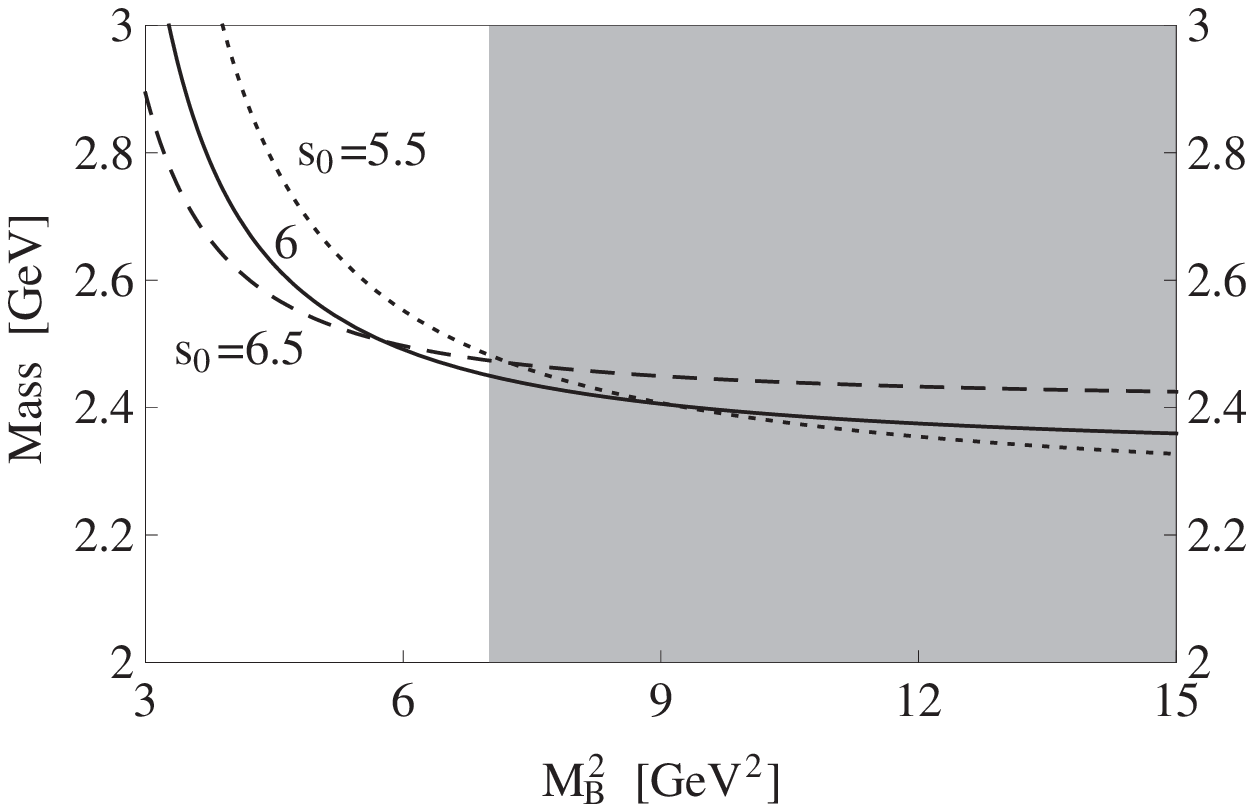}}
\centerline{\hspace{0.10in} {(a)} \hspace{2.85in}{ (b)}}
\caption{The variation of $M_{d^*}$ with respect to the threshold value $s_0$ and the Borel mass $M_B$. The current $J_2^{\alpha_1\alpha_2\alpha_3}$ is used here.}
\label{fig:current2}
\end{center}
\end{figure}

Similarly, we use the second current $J_2^{\alpha_1\alpha_2\alpha_3}$ to perform the QCD sum rule analyses. The results shown in Fig.~\ref{fig:current2} are quite similar to Fig.~\ref{fig:current1} obtained by using the first current $J_1^{\alpha_1\alpha_2\alpha_3}$. We obtain the mass of the $d^*(2380)$ to be
\begin{eqnarray}
M_{d^*} = 2.4\pm0.2 \mbox{ GeV} \, ,
\end{eqnarray}
and the decay constant $f_2$ to be:
\begin{eqnarray}
f_2 = (5.2\pm2.5) \times 10^{-5} \mbox{ GeV}^8 \, .
\end{eqnarray}

\section{Mixing of currents}
\label{sec:mixing}
In the previous section we used three single $\Delta-\Delta$ type dibaryon interpolating field, $J_a^{\alpha_1\alpha_2\alpha_3}$ ($a=1,2,3$), and performed the QCD sum rule analyses. Because the first and the third currents lead to identical sum rules, here we only study the mixing current constructed from the first and the second ones, which can be written as:
\begin{eqnarray}
J_{\rm mix}^{\alpha_1\alpha_2\alpha_3} = t_1 J_1^{\alpha_1\alpha_2\alpha_3} + t_2 J_2^{\alpha_1\alpha_2\alpha_3} \, ,
\label{eq:mixing}
\end{eqnarray}
where $t_1$ and $t_2$ are two real mixing parameters. %From Eqs.~(\ref{eq:currents1}) and (\ref{eq:currents2}) we can find that this current can be used generally to describe local dibaryon currents of color-singlet-color-singlet $\Delta-\Delta$ type.
Using this mixed current, we arrive at the following sum rule:
%
%%%%%%%%%%%%%%%%%%%%%%%%%%%%%%%%%%%%%%%%%%%%%%%%%%%%%%%%%%%%%%%%%%%%%%%%%%%%%%
\begin{eqnarray}
\label{eq:mixsumrule}
f^2_{\rm mix} e^{-M_{d^*}^2/M_B^2} &=& \Pi_{\rm mix}(s_0, M_B^2)
\\ \nonumber &=& \int^{s_0}_{0} e^{-s/M_B^2} ds \times \Big (
\frac{1}{8! 8! \pi ^{10}} s^7 \times \big ( {9\over44} t_1^2 - {9\over44} t_1t_2 + {45\over88}t_2^2 \big )
\\ \nonumber && - \frac{\langle g_s^2 GG \rangle}{ 8! 8! \pi^{10}} s^5 \times \big ( {343\over16} t_1^2 - {343\over16} t_1t_2 + {2989\over64}t_2^2 \big )
+ \frac{\langle \bar q q \rangle^2}{ 4! 4! \pi^6} s^4 \times \big ( {1\over14} t_1^2 - {17\over112} t_1t_2 + {13\over224}t_2^2 \big )
\\ \nonumber && + \frac{\langle \bar q q \rangle \langle g_s \bar q \sigma G q \rangle}{ 4! 4! \pi^6} s^3 \times \big ( {2\over3} t_1^2 - {17\over12} t_1t_2 + {13\over24}t_2^2 \big )
+ \frac{\langle g_s \bar q \sigma G q \rangle^2}{ 4! 4! \pi^6} s^2 \times \big ( {9\over10} t_1^2 - {153\over80} t_1t_2 + {117\over160}t_2^2 \big )
\\ \nonumber && - \frac{\langle g_s^2 GG \rangle \langle \bar q q \rangle^2}{4! 4! \pi^6} s^2 \times \big ( {7\over20} t_1^2 - {23\over40} t_1t_2 + {1\over5}t_2^2 \big )
+ \frac{\langle \bar q q \rangle^4}{\pi ^2} s \times \big ( {8\over27} t_1^2 - {17\over27} t_1t_2 + {13\over54}t_2^2 \big )
\\ \nonumber && - \frac{\langle g_s^2 GG \rangle \langle \bar q q \rangle \langle g_s \bar q \sigma G q \rangle}{ 4! 4! \pi^6} s \times \big ( {23 \over 24 } t_1^2 - {73\over48} t_1t_2 + {25\over48}t_2^2 \big )
\\ \nonumber && + \frac{\langle \bar q q \rangle^3 \langle g_s \bar q \sigma G q \rangle}{\pi^2} \times \big ( {4 \over 9} t_1^2 - {17\over18} t_1t_2 + {13\over36}t_2^2 \big )
- \frac{\langle g_s^2 GG \rangle \langle g_s \bar q \sigma G q \rangle^2}{ 4! 4! \pi^6} \times \big ( {3 \over 16} t_1^2 - {9\over32} t_1t_2 + {3\over32}t_2^2 \big )
\Big )
\\ \nonumber && - \frac{\langle g_s^2 GG \rangle \langle \bar q q \rangle^4}{\pi ^2} \times \big ( {1 \over 162} t_1^2 - {1\over162} t_1t_2 + {1\over648}t_2^2 \big ) \, .
\end{eqnarray}
%%%%%%%%%%%%%%%%%%%%%%%%%%%%%%%%%%%%%%%%%%%%%%%%%%%%%%%%%%%%%%%%%%%%%%%%%%%%%%
%
After performing numerical analysis, we find that as long as the Borel transformed two-point correlation function $\Pi_{\rm mix}(s_0, M_B^2)$ is positive in our working region $s_0$ around $5\sim7$ GeV$^2$, the results are also quite similar: we obtain the same mass of the $d^*(2380)$ of $M_{d^*} = 2.4\pm0.2$ GeV. Therefore, we arrive at the conclusion that the mixing of $\Delta-\Delta$ like local dibaryon currents does not change our previous conclusion significantly.

\section{Summary}
\label{sec:summary}

In summary, we have systematically constructed $I(J^P)=0(3^+)$ six-quark local interpolating currents without derivatives. We have presented arguments for selecting three $\Delta-\Delta$ like operators to perform QCD sum rule analyses and calculated the mass of the $d^*(2380)$. The OPE was calculated up to dimension 16 and we verified that its convergence is acceptable.The mass predictions are stable under variations in the two sum-rule parameters in the following regions
\begin{eqnarray}
7 \mbox{ GeV}^2 < M_B^2 < 15 \mbox{ GeV}^2 \, , 5.5 \mbox{ GeV}^2 < s_0 < 6.5 \mbox{ GeV}^2 \, ,
\end{eqnarray}
The sum-rule prediction of the $d^*(2380)$ mass is $M_{d^*} = 2.4\pm0.2$ GeV and its corresponding decay constant is $f_1 = f_3 = (5.5\pm2.7) \times 10^{-5}$ GeV$^8$ and $f_2 = (5.2\pm2.5) \times 10^{-5}$ GeV$^8$. Our result is consistent with the experimental value of the $d^*(2380)$ mass~\cite{Bashkanov:2008ih,Adlarson:2011bh,Adlarson:2012fe,Adlarson:2014pxj,Adlarson:2014ozl,Adlarson:2014tcn}. We also obtained a lower bound of  $M_{d^*} > 2.25$ GeV for the $d^*(2380)$ mass.
Considering the dibaryon current used in this paper is a $\Delta-\Delta$ like current, where two $\Delta$'s are both color singlets (see Eqs.~\eqref{eq:deltadelta1}-\eqref{eq:deltadelta3}), this result may suggest that the $\Delta-\Delta$ bound state has a binding energy less than 220 MeV.
To ensure that our conclusions are not significantly influenced by the choice of interpolating current,
 we mix these $\Delta-\Delta$ dibaryon operators, and calculate the two-point correlation using this mixed interpolating current and then perform the QCD sum rule analysis to extract the mass of the $d^*(2380)$ and its lower mass bound. The numerical results show that this
mixing effect does not change our previous conclusion for the $d^*(2380)$.

However, all the dibaryon currents discussed in this paper are local currents without derivatives. This makes it challenging to determine whether the $d^*(2380)$ is a $\Delta \Delta$ bound state or a six-quark state with hidden-color configurations, because we can always use the Fierz transformation to change a color-octet-color-octet structure to be a combination of color-singlet-color-singlet structures (see Eqs.~(\ref{eq:color1}) and (\ref{eq:color2})), and vice versa. To discriminate between them in the QCD sum rule, one can use a non-local current, such as
\begin{eqnarray}
J_{\alpha_1\alpha_2\alpha_3}(x,y) = \mathcal{S} [(\Delta^{++}_{1\alpha_1}(x))^T C \gamma_{\alpha_3} \Delta^{-}_{1\alpha_2}(y)] \, ,
\end{eqnarray}
where one $\Delta$ state is at location $x$ and the other at $y$. Other areas of future investigation could include the use of currents containing
derivative operators to perform QCD sum rule analyses.

\section*{Acknowledgments}

We thank Dian-Yong Chen and Hai-Qing Zhou for useful discussion and information on the polarization tensor $\eta_{\alpha_1\alpha_2\alpha_3}$. This work is supported by the National Natural Science Foundation of China under Grant No.11205011, 11475015 and 11261130311. Wei Chen and T. G. Steele are supported by the Natural Sciences and Engineering
Research Council of Canada (NSERC).

\end{document}